# Optimizing Emotion Recognition with Wearable Sensor Data: Unveiling Patterns in Body Movements and Heart Rate through Random Forest Hyperparameter Tuning

Zikri Kholifah Nur\*, Rifki Wijaya, Gia Septiana Wulandari[3,]

School of Computing, Informatics, Telkom University, Bandung, Indonesia
Email: \*zikrikn@student.telkomuniversity.ac.id, rifkiwijaya@telkomuniversity.ac.id, giaseptiana@telkomuniversity.ac.id
Correspondence Author Email: zikrikn@student.telkomuniversity.ac.id

**Abstract** − This research delves into the utilization of smartwatch sensor data and heart rate monitoring to discern individual emotions based on body movement and heart rate. Emotions play a pivotal role in human life, influencing mental well-being, quality of life, and even physical and physiological responses. The data were sourced from prior research by Juan C. Quiroz, PhD. The study enlisted 50 participants who donned smartwatches and heart rate monitors while completing a 250-meter walk. Emotions were induced through both audio-visual and audio stimuli, with participants' emotional states evaluated using the PANAS questionnaire. The study scrutinized three scenarios: viewing a movie before walking, listening to music before walking, and listening to music while walking. Personal baselines were established using DummyClassifier with the 'most_frequent' strategy from the sklearn library, and various models, including Logistic Regression and Random Forest, were employed to gauge the impacts of these activities. Notably, a novel approach was undertaken by incorporating hyperparameter tuning to the Random Forest model using RandomizedSearchCV. The outcomes showcased substantial enhancements with hyperparameter tuning in the Random Forest model, yielding mean accuracies of 86.63% for happy vs. sad and 76.33% for happy vs. neutral vs. sad.

## 1. INTRODUCTION

Emotion, an integral part of human experience, influences various aspects of daily life, from mental well-being and social interactions to overall quality of life [1], [2]. The complexity of modern life necessitates innovative methods for comprehensive assessment and management of emotions [3]. Wearable technology, such as smartwatches and fitness trackers, offers new avenues for emotion research, enabling real-time monitoring and analysis of emotional experiences in naturalistic settings [4], [5]. The convergence of psychology and technology holds the potential to advance our understanding of emotional phenomena and create effective programs to enhance mental health and well-being [6].

Previous studies have explored the use of wearable technology for emotion detection based on physiological responses and body movements [7], [8], [9], [10]. A notable study by Juan C. Quiroz, PhD, investigated the potential of smartwatch sensors to recognize emotional states by monitoring physiological signals such as heart rate and movement data [10]. Quiroz's research demonstrated that it is feasible to detect emotions using data collected from wearable devices [10]. The research employed a mixed design methodology and leveraged machine learning techniques to classify emotional states, consistently achieving median accuracies exceeding 78% across all conditions for the binary classification of happiness versus sadness [10]. However, the study could benefit from further exploration to potentially enhance the accuracy and robustness of the emotion recognition models [11].

Building on the foundational work of Quiroz, our study introduces a novel approach to emotion detection that leverages data from multiple sensors embedded in smartwatches and heart rate monitors. The study involves 50 participants who wore these devices while walking 250 meters. Emotional states were induced using audio-visual and audio stimuli, and the participants' emotional conditions were evaluated using the Positive and Negative Affect Schedule (PANAS) questionnaire [12]. Unlike Quiroz's study, which used a static set of parameters for machine learning models, our research implements the Random Forest algorithm with hyperparameter tuning using RandomizedSearchCV [13]. The Random Forest algorithm is an ensemble learning method that constructs multiple decision trees and merges their results to improve predictive performance and control overfitting [14]. By tuning the hyperparameters of the Random Forest model using RandomizedSearchCV, we systematically sample different combinations of hyperparameters and evaluate their performance using cross-validation, aiming to find the optimal configuration that maximizes the model's predictive accuracy and generalizability [15].

Hyperparameter tuning is crucial as it involves selecting the best combination of parameters such as the number of trees in the forest, the maximum depth of each tree, and the minimum number of samples required to split a node in a Random Forest model [16]. This process enhances the model's ability to generalize from the training data to unseen data, thereby improving its accuracy and reliability in emotion detection tasks [17].

Our study builds on the existing body of research by specifically addressing the limitations of Quiroz's work regarding model optimization. By incorporating hyperparameter tuning with the Random Forest algorithm, we introduce a significant advancement that offers a promising solution to enhance the accuracy and robustness in emotion recognition [18]. This enhancement is expected to yield better predictive performance and reliability, thereby advancing the field of emotion detection using wearable technology [19].

By combining physiological signals and movement data, this study aims to uncover unique patterns associated with different emotional states [20]. This comprehensive approach not only enhances the detection capabilities but also contributes to a deeper understanding of the interplay between physiological responses and emotions. Furthermore, the use of advanced machine learning techniques ensures that our findings are both reliable and applicable to a wide range of real-world scenarios [21].

Our research strives to advance the field of emotion detection by introducing a robust and accurate method that leverages the strengths of wearable technology and sophisticated machine learning algorithms. The use of Random Forest with hyperparameter tuning holds significant promise for improving the detection and analysis of emotions, ultimately contributing to better mental health monitoring and personalized user experiences [22]. This study represents a step forward in the quest for non-invasive, real-time emotion detection systems that can be seamlessly integrated into daily life, providing valuable insights for both researchers and practitioners in the field of affective computing.

## 2. RESEARCH METHODOLOGY

**2.1 System Architecture**

Previously, the system architecture was developed by Juancq and uploaded to their GitHub repository. The process involved data collection, preprocessing (including creating walking data and performing feature extraction), and the implementation of personal models. These models included a baseline using DummyClassifier with the 'most_frequent' strategy, Logistic Regression, and Random Forest. However, a novel addition to this system architecture is the incorporation of Random Forest Hyperparameter tuning using the Randomized Search method. Subsequently, evaluations were conducted to assess the performance of each model. The process flow is illustrated in the flowchart in Figure 1.

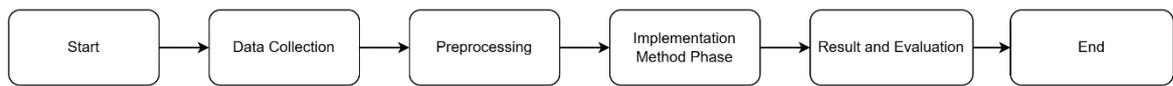

**Figure 1.** System architecture flow from data collection, preprocessing, implementation method phase, until result and evaluation

**2.1.1 Data Collection**

The data collection process involved enlisting 50 young adult volunteers from a university campus in North-West UK. Predominantly female, comprising 43 individuals, the participants had an average age of 23.18 years. As payment for taking part in the study, each volunteer earned 7 euros. It's important to note that all participants reported no visual or hearing impairments and were capable of walking unaided. Recruitment was carried out through notices on bulletin boards and word of mouth, ensuring a representation of healthy and active campus individuals.

This mixed-design study aimed to explore emotions, including happy, sad, and neutral states, induced through audio-visual stimuli, consisting of movie clips, and audio stimuli, including music clips. Throughout the experiment, participants wore smartwatches on their left wrists and heart-rate straps fastened around their chests. The heart-rate straps, utilizing the Polar H7 sensor, facilitated the collection of PANAS (Positive Affect and Negative Affect) scores. The experiment was conducted three times to adequately cover each emotional state. The data were obtained from the Juancq/emotion-recognition-smartwatch repository on GitHub. The smartwatches, specifically Samsung Gear 2, were equipped with tri-axial accelerometers and tri-axial gyroscopes, capturing participants' movements. These sensors were worn by participants throughout the experiment duration. The experimental conditions included: (1) participants watching movie clips before walking, (2) participants listening to music clips before walking, and (3) participants listening to music while walking. Heart rate monitoring was facilitated by the Polar M400 watch worn by the experimenter, who monitored the data in real-time. After collecting the data we will get the CSV files and then we continue to preprocessing data. Figure 2 illustrates the flowchart of the data collection process, detailing each step from data collection to data preprocessing.

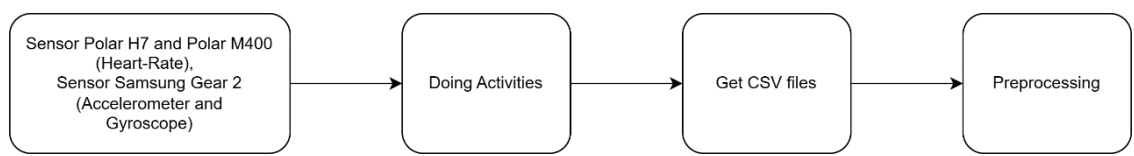

**Figure 2.** Flowchart of Data Collection

Table 1 shows an overview of the user study encoding, which encompasses participant ID, condition, age, sex, and the start and end times for each condition. The participant ID serves as a unique identifier for each individual involved in the study, facilitating data tracking and analysis. The condition refers to the specific experimental setup or treatment administered to participants, guiding the tasks or stimuli they encountered. Age provides demographic insight by indicating the age of each participant during the study, while sex denotes their gender. These demographic factors contribute to understanding potential variations in responses. Additionally, the start and end times for each condition denote the duration of experimental phases, aiding in temporal synchronization and analysis.

**Table 1.** The user-related data includes participant id, condition, age, sex, and the start and end times for each condition

| PARTICIPANT ID | CONDITION | AGE | SEX | START W | END W | START W | END W | START W | END W |
|---|---|---|---|---|---|---|---|---|---|
| EW2 | Mo-SNH | 23 | F | 13.16.29 | 13.19.41 | 13.25.17 | 13.28.30 | 13.32.27 | 13.35.34 |
| EW3 | Mo-SNH | 37 | F | 14.29.09 | 14.32.21 | 14.38.01 | 14.41.14 | 14.44.51 | 14.47.56 |
| EW4 | Mo-HNS | 28 | F | 11.28.26 | 11.31.06 | 11.36.03 | 11.38.40 | 11.43.14 | 11.45.52 |
| EW5 | Mu-SNH | 31 | M | 16.21.16 | 16.24.49 | 16.30.00 | 16.33.41 | 16.38.40 | 16.42.14 |
| EW6 | Mu-HNS | 27 | F | 19.18.54 | 19.22.06 | 19.27.21 | 19.30.40 | 19.35.53 | 19.39.17 |

Table 2 shows an overview of the raw data collected in CSV format, including various sensor readings such as acceleration (ax, ay, az), gravity-compensated acceleration (ax_g, ay_g, az_g), rotation (rot_x, rot_y, rot_z), and heart rate. These data, captured from wearable devices like smartwatches and heart rate monitors, offer valuable insights into participants' physiological responses and movements throughout the study, facilitating analysis of the relationship between physiological signals and emotional states or behaviors.

**Table 2.** The data obtained at this stage is a CSV files containing ax, ay, az, ax_g, ay_g, az_g, rot_x, rot_y, rot_z, and heart for each users

| time | ax | ay | az | ax_g | ay_g | az_g | rot_x | rot_y | rot_z | heart |
|---|---|---|---|---|---|---|---|---|---|---|
| 11:22:31:148 | -0.25 | 0.55 | 5.88 | -0.39 | 0.86 | 9.18 | 102.76 | 38.85 | 46.13 | 74 |
| 11:22:31:178 | -0.59 | 0.26 | 4.91 | -0.88 | 0.63 | 9.44 | -21.00 | -9.31 | -0.21 | 74 |
| 11:22:31:218 | -0.24 | 0.81 | 3.75 | -0.59 | 1.38 | 9.22 | -27.58 | -25.90 | -3.43 | 74 |
| 11:22:31:258 | -0.34 | 0.24 | 3.82 | -0.78 | 0.88 | 10.25 | -28.98 | -13.72 | -16.10 | 74 |

## 2.1.2 Preprocessing

Preprocessing data involves several steps aimed at preparing the raw data for further analysis and modeling. The process typically starts with reading the encoding file, which contains information about each participant, such as their ID, age, sex, and the start and end times for each condition. The participant ID is then matched with the corresponding data file containing accelerometer, gyroscope, and heart rate data. The data is read from the matched file, and the time data along with the sensor data is extracted. The start and stop times are determined to identify the duration of each activity. Emotion and condition columns are created based on the experimental conditions, and the column order is arranged accordingly. The processed walking data is then saved to a CSV file for further analysis. Following this preprocessing step, the process proceeds to feature extraction. Figure 3 visually summarizes the workflow for generating and preparing walking data, culminating in the feature extraction stage.

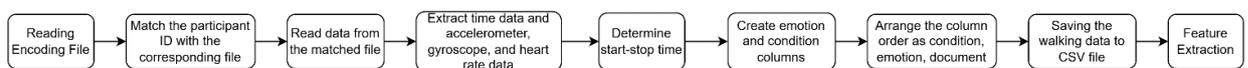

**Figure 3.** The flowchart for generating walking data is followed by the subsequent step, which is feature extraction

In the feature extraction process, several variables are initialized, including window size, overlap, input files, output directory, delimiter, and frequency rate. The data is read from the input file, which includes information about the condition, emotion, and sensor data. Emotions are extracted, and frames are created based on emotion indices to segment the data. A filter is applied to eliminate noise in the accelerometer data, ensuring the quality of the extracted features. Features are then extracted from the segmented data, resulting in a list of 107 features representing various aspects of the walking patterns and physiological responses. Once the feature extraction is complete, the process proceeds to the implementation method phase, where machine learning models are trained and evaluated using the extracted features. Figure 4 picks up where Figure 3 leaves off, illustrating the subsequent steps from feature extraction through to the implementation of machine learning models.

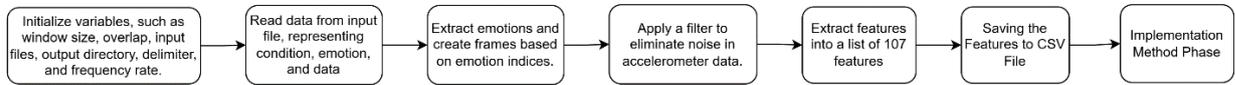

**Figure 4.** The flowchart for feature extraction is followed by the implementation method phase

### 2.1.3 Implementation Method Phase

In the Implementation Method Phase, the chosen models or methods are implemented according to the predetermined approach. For this case, the models used include DummyClassifier with the strategy 'most_frequent' as the baseline, logistic regression, random forest, and random forest with hyperparameter tuning. Each phase for the model follows a similar process: Initialization, where the model is initialized, including setting any necessary parameters or configurations. Next is Model Training with Training Data, where the model is trained using the training data to learn the underlying patterns or relationships in the data. Following that, Cluster Label Prediction occurs, where once trained, the model is used to predict cluster labels or outcomes for new or unseen data. There is an exception for Random Forest with Hyperparameter Tuning Implementation, which includes an additional step: Hyperparameter Tuning Phase. In this step, the hyperparameters of the random forest model are tuned to optimize its performance. This phase occurs before the cluster label prediction step. Figure 5 illustrates the flowchart of the Implementation Method Phase, showing the sequence of steps from model initialization to the final prediction, including details on how the models are configured and evaluated.

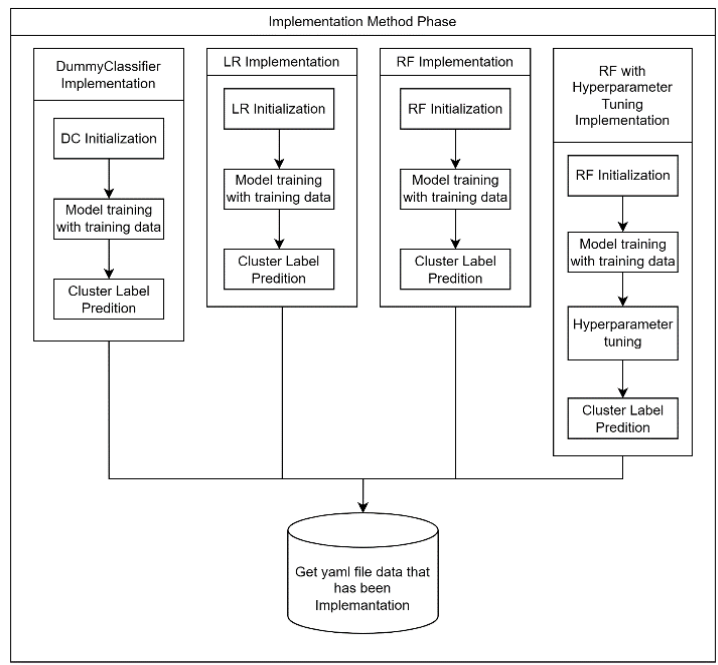

**Figure 5.** This depicts the flowchart for each model in the Implementation Method Phase, where the output is in YAML format. The YAML file generated can be applied to assess how well the models being employed perform. This evaluation helps assess how well each model performs in predicting cluster labels or outcomes based on the input data.

#### 2.1.3.1 DummyClassifier (Scikit-learn) as a baseline model

The DummyClassifier from sklearn serves as a rudimentary classification model used as a baseline for comparison with more sophisticated classifiers. It operates by making predictions independent of input features (X), essentially ignoring the feature values during prediction. The primary purpose of the DummyClassifier is to provide a reference point for evaluating the performance of more complex models. In this study, the "most_frequent" strategy is employed, wherein the predict method always returns the class label that is most frequently observed in the training data (y) during the fit method invocation. Similarly, the predict_proba method returns the empirical class distribution of y, representing the prior class distribution. Essentially, this model consistently predicts the class label with the highest frequency in the training data and its probability distribution follows the empirical prior class distribution. This strategy offers a simplistic yet useful comparison baseline and serves as a straightforward "dumb" model. Comparing the performance of more sophisticated models against this basic baseline allows researchers to gauge the effectiveness and added value of complexity in their models. If the performance of the DummyClassifier is comparable to the best-performing model, it suggests the need for further investigation into the data or model architecture. Figure 6 illustrates the implementation of the 'most_frequent' strategy in the DummyClassifier, showing how the model predicts using the most frequently occurring class label in the training data.

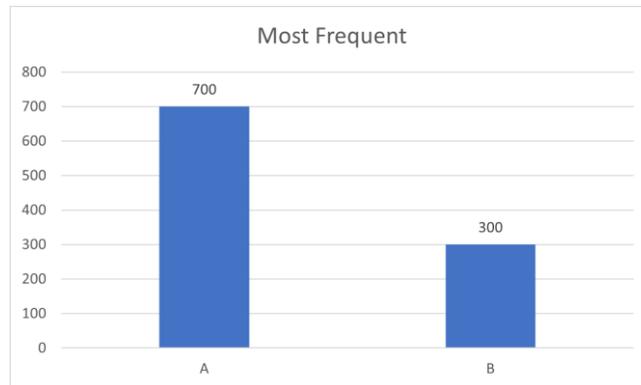

**Figure 6.** A bar chart illustrating the implementation of the 'most_frequent' strategy in the DummyClassifier as a baseline.

### 2.1.3.2 Logistic Regression

Logistic regression is a statistical method employed to predict the probability of a binary outcome based on one or more predictor variables. It models the relationship between the predictors and the log-odds of the event, utilizing a sigmoid function to transform the linear prediction into a probability between 0 and 1 [23]. The model's parameters are estimated using maximum likelihood estimation, enabling the interpretation of coefficients in terms of the impact of predictors on the outcome's likelihood. While inherently designed for binary classification, logistic regression can be extended to handle multi-class problems through techniques like one-vs-rest or multinomial logistic regression, broadening its applicability across diverse classification tasks. Figure 7 illustrates the fundamental concept of Logistic Regression. It showcases the characteristic S-shaped curve, also known as the sigmoid function, which represents the model's output - the predicted probability of the binary outcome (Y) given the input variable (X). The curve demonstrates how the predicted probabilities lie between 0 and 1, smoothly transitioning from low to high probabilities as the value of X increases. The red dots superimposed on the graph might represent the actual observed data points, highlighting how the logistic regression model fits the data and estimates the probability of belonging to a particular class. The position of these points in relation to the curve visually indicates the model's accuracy in classifying the data.

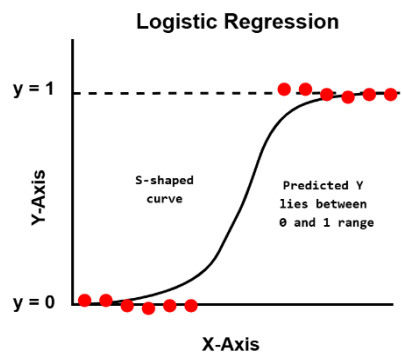

**Figure 7.** Visualization of Logistic Regression

### 2.1.3.3 Random Forest

In machine learning, the Random Forest algorithm is a potent ensemble learning method that's frequently used to solve issues with both regression and classification. It belongs to the family of tree-based models and is known for its robustness and versatility across various domains. A collection of decision trees, each trained on a random subset of the training data and features, is the fundamental building block of Random Forest. This approach of randomization during training lessens the likelihood of overfitting and enhances the model's ability to generalize by decorrelationing the individual trees [24].

Individually, each decision tree in a Random Forest is constructed by recursively partitioning the feature space into regions or leaves based on feature values. The splitting criteria aim to minimize impurity or maximize information gain at each node. However, unlike traditional decision trees, Random Forest introduces additional randomness by randomly sampling both the data and features at each split.

A forest of trees is created during the training phase as a consequence of the independent growth of several decision trees. Each tree in the forest independently forecasts the target variable for regression tasks or the input data for classification tasks in order to generate predictions. In classification, the final prediction is typically determined by a majority vote among the trees, while in regression, it is often the average of the predictions from all trees. Mathematically, the prediction process of a Random Forest can be described as follows:

For classification:

$$\hat{y} = mode(y1, y2, \ldots, y_n) \quad (2)$$

where $\hat{y}$ represents the predicted class label, and $y1, y2, \ldots, y_n$ are the class labels predicted by each decision tree in the forest.

Random Forest offers several advantages over other machine learning algorithms. It is robust to overfitting, thanks to the ensemble nature of the model and the inherent randomness in tree construction. Moreover, it is resistant to noise in the data and capable of handling high-dimensional feature spaces efficiently. Due to its high performance and ease of use, Random Forest finds applications in various domains, including healthcare, bioinformatics, and natural language processing. Figure 8 illustrates the step-by-step process of Random Forest Classification, from dataset division to final prediction through majority voting of multiple decision trees.

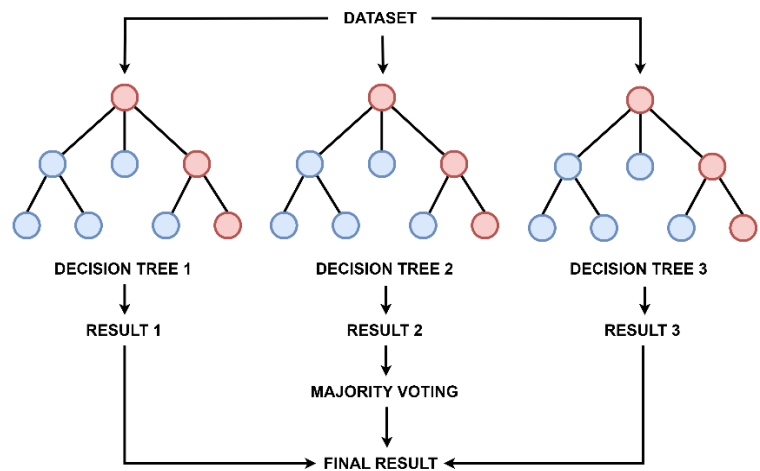

**Figure 8.** Visualization of the Process of Random Forest Classification

### 2.1.3.4 Random Forest with Hyperparameter Tuning (Using RandomSearch)

Random Forest with Hyperparameter Tuning using Random Search is an advanced technique aimed at optimizing the performance of a Random Forest model by fine-tuning its hyperparameters. Hyperparameters are configuration settings that control the behavior of the model during the training process, things like the total number of trees in the forest, each tree's maximum depth, and the least amount of samples needed to divide a node [25].

The Random Search method involves systematically sampling hyperparameter combinations from a predefined search space and evaluating the performance of each combination using cross-validation. Unlike grid search, which exhaustively evaluates all possible combinations, random search randomly selects a subset of hyperparameter combinations to evaluate, making it more computationally efficient, especially for large search spaces. During the hyperparameter tuning process, the Random Forest model is trained multiple times using different combinations of hyperparameters. Each trained model is then evaluated on a validation set using a performance metric such as accuracy, precision, recall, or F1 score. The hyperparameter combination that yields the best performance metric on the validation set is selected as the optimal configuration for the Random Forest model. The advantage of using Random Search for hyperparameter tuning is that it allows for a more thorough exploration of the hyperparameter space compared to manual tuning or simple heuristics. By randomly sampling hyperparameter combinations, Random Search can efficiently discover promising regions of the search space, even in high-dimensional or non-convex spaces.

Random Forest with Hyperparameter Tuning using Random Search is a powerful technique for optimizing the performance of Random Forest models, leading to improved accuracy and generalization on unseen data. It is commonly used in machine learning projects to achieve state-of-the-art performance in classification and regression tasks.

## 3. RESULT AND DISCUSSION

### 3.1 Results of Data Preprocessing

The preprocessing stage encompassed two main tasks: generating walking data and extracting features from the raw sensor data. This processed data serves as the foundation for the subsequent implementation phase.

#### 3.1.1 Generated Walking Data

The walking data was generated through the processing of raw data collected under various conditions, including watching movie clips, listening to music clips, and walking while listening to music. Each table within the walking data section presents accelerometer and gyroscope readings, heart rate data, as well as associated emotions and conditions for each participant during different activities.

**Table 3.** Watching movie clips walking data

| condition | emotion | ax | ay | az | rot_x | rot_y | rot_z | heart |
|---|---|---|---|---|---|---|---|---|
| 0 | 1 | -2.20 | -0.26 | 1.32 | -91.28 | 8.89 | -125.51 | 81 |
| 0 | 1 | -2.68 | -2.11 | 0.84 | -122.85 | -6.93 | -112.77 | 81 |
| 0 | 1 | -2.35 | -2.14 | 0.84 | -35.14 | -10.43 | -90.65 | 81 |
| 0 | 1 | -2.46 | -1.73 | 0.47 | -16.10 | -18.55 | -54.32 | 81 |
| 0 | 1 | -2.50 | -1.50 | -0.28 | -31.15 | -40.74 | -29.89 | 81 |

**Table 4.** Listening to music clips walking data

| condition | emotion | ax | ay | az | rot_x | rot_y | rot_z | heart |
|---|---|---|---|---|---|---|---|---|
| 1 | 0 | 0.18 | -0.08 | -1.02 | 10.08 | 14.91 | -60.48 | 117 |
| 1 | 0 | -0.26 | 1.32 | -0.40 | 77.35 | 24.57 | -53.76 | 117 |
| 1 | 0 | 0.66 | 1.52 | 0.26 | 51.45 | 3.08 | -60.48 | 117 |
| 1 | 0 | 3.51 | 1.42 | 0.21 | 32.62 | 0.28 | -52.64 | 117 |
| 1 | 0 | 3.54 | 0.71 | 0.65 | 33.95 | 14.70 | -32.76 | 117 |

**Table 5.** Listening to music while walking data

| condition | emotion | ax | ay | az | rot_x | rot_y | rot_z | heart |
|---|---|---|---|---|---|---|---|---|
| 2 | -1 | 0.39 | 0.33 | -0.18 | -5.88 | 34.72 | 27.72 | 85 |
| 2 | -1 | 0.05 | -0.33 | 0.96 | 7.28 | 27.02 | 29.89 | 85 |
| 2 | -1 | 0.78 | 0.01 | -0.69 | 6.16 | 36.47 | 26.53 | 85 |
| 2 | -1 | 0.46 | 0.12 | 1.10 | 12.74 | 36.05 | 32.06 | 85 |
| 2 | -1 | 0.79 | 0.20 | -1.21 | 20.37 | 44.80 | 26.53 | 85 |

Tables 3, 4, and 5 present an overview of the processed data transformed into walking data. Each row in the tables represents a data point recorded during the experiments. The "condition" column indicates the experimental conditions followed by the participants, where 0 denotes watching movie clips before walking, 1 denotes listening to music before walking, and 2 denotes listening to music while walking. The "emotion" column represents the expected emotions elicited in the participants after being exposed to stimuli, with 1 for happy, 0 for neutral, and -1 for sad. Columns "ax," "ay," and "az" represent the acceleration values along the x, y, and z axes recorded by the smartwatch, while "rot_x," "rot_y," and "rot_z" denote the rotation values along the same axes. Finally, the "heart" column indicates the participants' heart rates recorded by the heart rate monitor. These data are then utilized to extract relevant features such as mean, standard deviation, and maximum values from each sensor, which are subsequently used as inputs to train machine learning models for emotion recognition based on sensor data.

#### 3.1.2 Feature Extraction

Moving on to feature extraction, this stage focused on identifying essential patterns and characteristics from the walking data. Although a total of 107 features were derived, we will highlight only 12. These features encompass various aspects of participants' movements and physiological responses, including the standard deviations of accelerometer and gyroscope readings, angles, magnetic field strength, and heart rate. Each table in the feature extraction section presents a subset of these features, along with the corresponding emotions and conditions for each participant during different activities. These extracted features serve as input variables for the subsequent modeling phase, facilitating the development of predictive models to classify emotional states based on sensor data.

**Table 6.** Watching movie clips extraction data

| acc_x_std | acc_y_std | acc_z_std | gyro_x_std | gyro_y_std | gyro_z_std | angle_x | angle_y | angle_z | mag | heart | emotion |
|---|---|---|---|---|---|---|---|---|---|---|---|
| 1.96 | 1.36 | 1.13 | 71.05 | 45.14 | 68.82 | 1.09 | 2.11 | 2.38 | 0.67 | 81 | 1 |
| 1.31 | 1.39 | 1.01 | 96.83 | 55.03 | 49.57 | 0.96 | 1.98 | 2.37 | 0.73 | 81 | 1 |
| 0.95 | 1.58 | 1.45 | 134.85 | 67.30 | 77.36 | 0.78 | 2.33 | 1.73 | 0.89 | 81 | 1 |
| 1.13 | 1.16 | 1.40 | 133.53 | 77.58 | 116.27 | 0.85 | 2.19 | 1.07 | 0.87 | 81 | 1 |
| 1.15 | 1.79 | 1.08 | 116.86 | 46.59 | 134.20 | 1.07 | 0.53 | 1.74 | 1.00 | 81 | 1 |

**Table 7.** Listening to music clips extraction data

| acc_x_std | acc_y_std | acc_z_std | gyro_x_std | gyro_y_std | gyro_z_std | angle_x | angle_y | angle_z | mag | heart | emotion |
|---|---|---|---|---|---|---|---|---|---|---|---|
| 1.68 | 0.71 | 0.60 | 47.20 | 20.87 | 42.39 | 2.42 | 2.29 | 1.60 | 0.75 | 119 | 0 |
| 1.95 | 0.71 | 0.46 | 42.74 | 20.82 | 50.60 | 1.71 | 1.58 | 3.00 | 0.86 | 119 | 0 |
| 2.07 | 0.80 | 0.31 | 41.49 | 18.97 | 50.33 | 0.17 | 1.55 | 1.40 | 0.82 | 119 | 0 |
| 1.78 | 0.75 | 0.29 | 41.71 | 16.75 | 51.45 | 0.33 | 1.24 | 1.57 | 0.77 | 119 | 0 |
| 1.67 | 0.76 | 0.43 | 47.10 | 21.41 | 42.73 | 0.26 | 1.50 | 1.33 | 0.80 | 120 | 0 |

**Table 8.** Listening to music while walking extraction data

| acc_x_std | acc_y_std | acc_z_std | gyro_x_std | gyro_y_std | gyro_z_std | angle_x | angle_y | angle_z | mag | heart | emotion |
|---|---|---|---|---|---|---|---|---|---|---|---|
| 0.23 | 0.46 | 0.41 | 11.11 | 4.38 | 8.18 | 2.15 | 2.56 | 1.49 | 0.39 | 108 | -1 |
| 0.20 | 0.16 | 0.16 | 12.53 | 5.57 | 9.21 | 2.11 | 1.10 | 2.39 | 0.20 | 109 | -1 |
| 0.31 | 0.36 | 0.44 | 11.70 | 11.53 | 11.07 | 1.99 | 0.42 | 1.62 | 0.36 | 109 | -1 |
| 0.35 | 0.34 | 0.40 | 13.46 | 11.92 | 7.68 | 1.73 | 0.41 | 1.20 | 0.33 | 109 | -1 |
| 0.51 | 0.22 | 0.39 | 17.28 | 15.50 | 8.75 | 1.56 | 0.03 | 1.55 | 0.21 | 109 | -1 |

Tables 6, 7, and 8 display an overview of features extracted from the walking data for each condition, with 12 columns highlighted here. These include the standard deviation of acceleration (acc_x_std, acc_y_std, acc_z_std) and gyroscope (gyro_x_std, gyro_y_std, gyro_z_std) values along the x, y, and z axes, indicating movement variability. Additionally, the angles between the mean acceleration vector and the x, y, and z axes (angle_x, angle_y, angle_z) offer information on device orientation during movement. The magnitude of acceleration (mag) is included to show overall acceleration variability. Lastly, participants' heart rates (heart) and the associated emotion labels (emotion) for each measurement are provided. These emotion labels are classified by the machine learning model, with 1 representing happy, 0 representing neutral, and -1 representing sad. Each row in these tables represents a data point with the extracted features and corresponding emotion label. These features are crucial inputs for the machine learning model to classify and predict emotions based on sensor data patterns.

### 3.2 Outcomes of Implementation and Analysis

In this section, the performance of various machine learning models, including the baseline, Logistic Regression, Random Forest, and Random Forest with Hyperparameter Tuning, was evaluated for distinguishing between happy and sad conditions, as well as between happy, sad, and neutral conditions.

#### 3.2.1 Happy vs. Sad

In the implementation phase, the performance of various machine learning models was assessed separately to distinguish between happy and sad conditions, as well as between happy, sad, and neutral conditions. For the happy vs. sad comparison, Logistic Regression, Random Forest, and Random Forest with Hyperparameter Tuning were evaluated. The results, showcased in Table 9 and the box plots in Figure 9, provide insights into the accuracy, area under the curve (AUC), F1 score, user lift, and p-values for each model across different conditions. Particularly, Random Forest with hyperparameter tuning consistently outperformed the baseline, Logistic Regression, and default Random Forest models in terms of accuracy and predictive performance for discriminating between happy and sad conditions. This suggests the robustness and efficacy of Random Forest models, especially when optimized through hyperparameter tuning, for classifying emotional states based on wearable sensor data.

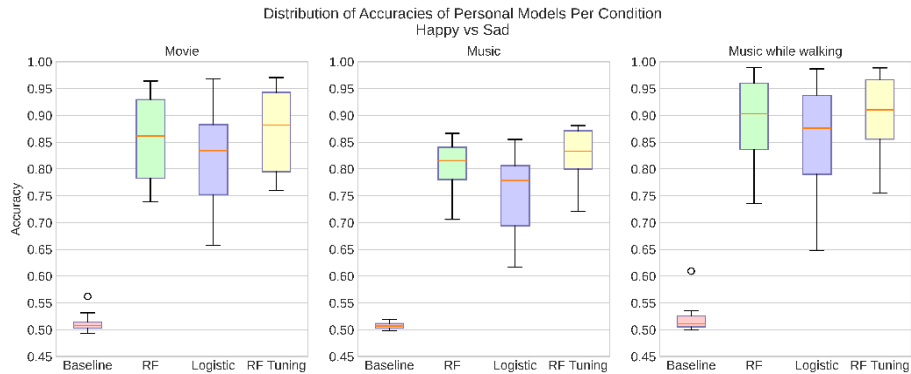

**Figure 9.** Boxplot of accuracies for distinguishing happy vs. sad conditions

**Table 9.** Average accuracy lift for happy and sad conditions across each condition

| Condition 1: Watch movie then walk | AUC | F1 | Accuracy | User Lift | P-value |
|---|---|---|---|---|---|
| Model | | | | | |
| Baseline | 0.500 (0.000) | 0.347 (0.018) | 0.512 (0.016) | | |
| Logistic Regression | 0.877 (0.084) | 0.818 (0.089) | 0.818 (0.089) | 0.306 | 0.000 |
| Random Forest | 0.922 (0.059) | 0.852 (0.074) | 0.853 (0.074) | 0.341 | 0.000 |
| Random Forest with Hyperparameter Tuning | 0.934 (0.054) | 0.871 (0.073) | 0.871 (0.072) | 0.359 | 0.000 |
| Condition 2: Listen to music then walk | AUC | F1 | Accuracy | User Lift | P-value |
| Model | | | | | |
| Baseline | 0.500 (0.000) | 0.342 (0.007) | 0.508 (0.006) | | |
| Logistic Regression | 0.813 (0.081) | 0.748 (0.072) | 0.749 (0.071) | 0.241 | 0.000 |
| Random Forest | 0.887 (0.046) | 0.807 (0.046) | 0.808 (0.045) | 0.300 | 0.000 |
| Random Forest with Hyperparameter Tuning | 0.905 (0.043) | 0.827 (0.047) | 0.827 (0.046) | 0.320 | 0.000 |

| Condition 3: Listen to music while walk | AUC | F1 | Accuracy | User Lift | P-value |
|---|---|---|---|---|---|
| Model | | | | | |
| Baseline | 0.500 (0.000) | 0.356 (0.031) | 0.520 (0.027) | | |
| Logistic Regression | 0.901 (0.095) | 0.849 (0.107) | 0.850 (0.107) | 0.329 | 0.000 |
| Random Forest | 0.948 (0.057) | 0.890 (0.081) | 0.891 (0.080) | 0.371 | 0.000 |
| Random Forest with Hyperparameter Tuning | 0.956 (0.048) | 0.901 (0.073) | 0.901 (0.072) | 0.381 | 0.000 |

### 3.2.2 Happy vs. Neutral vs. Sad

Similarly, for the comparison between happy, neutral, and sad conditions, Logistic Regression, Random Forest, and Random Forest with Hyperparameter Tuning were utilized, and their performance metrics were evaluated. The results, displayed in Table 10 and the box plots in Figure 10, demonstrate the accuracy, AUC, F1 score, user lift, and p-values for each model across different conditions. Once again, Random Forest models, particularly those optimized through hyperparameter tuning, consistently outperformed the baseline, Logistic Regression, and default Random Forest models in terms of accuracy and predictive performance. This underscores the superiority of Random Forest models in distinguishing between happy, sad, and neutral emotional states based on wearable sensor data, especially the Random Forest model with hyperparameter tuning.

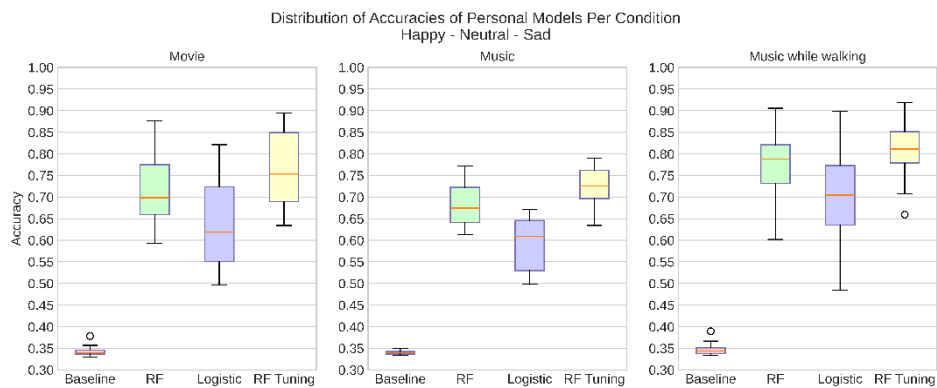

**Figure 10** Boxplot of accuracies for distinguishing happy vs. neutral vs. sad conditions

**Table 10** Average accuracy lift for happy, sad, and neutral conditions across each condition

| Condition 1: Watch movie then walk | AUC | F1 | Accuracy | User Lift | P-value |
|---|---|---|---|---|---|
| Model | | | | | |
| Baseline | 0.500 (0.000) | 0.175 (0.010) | 0.343 (0.011) | | |
| Logistic Regression | 0.798 (0.082) | 0.633 (0.104) | 0.635 (0.103) | 0.292 | 0.000 |
| Random Forest | 0.878 (0.061) | 0.723 (0.090) | 0.724 (0.090) | 0.381 | 0.000 |
| Random Forest with Hyperparameter Tuning | 0.903 (0.055) | 0.759 (0.085) | 0.760 (0.085) | 0.417 | 0.000 |
| Condition 2: Listen to music then walk | AUC | F1 | Accuracy | User Lift | P-value |
| Model | | | | | |
| Baseline | 0.500 (0.000) | 0.173 (0.004) | 0.340 (0.004) | | |
| Logistic Regression | 0.763 (0.058) | 0.590 (0.062) | 0.592 (0.061) | 0.252 | 0.000 |
| Random Forest | 0.854 (0.038) | 0.683 (0.049) | 0.685 (0.049) | 0.345 | 0.000 |

| | AUC | F1 | Accuracy | User Lift | P-value |
|---|---|---|---|---|---|
| Random Forest with Hyperparameter Tuning | 0.881 (0.035) | 0.720 (0.048) | 0.721 (0.047) | 0.381 | 0.000 |
| Condition 3: Listen to music while walk Model | AUC | F1 | Accuracy | User Lift | P-value |
| Baseline | 0.500 (0.000) | 0.179 (0.014) | 0.348 (0.015) | | |
| Logistic Regression | 0.853 (0.084) | 0.710 (0.115) | 0.711 (0.114) | 0.363 | 0.000 |
| Random Forest | 0.916 (0.053) | 0.781 (0.085) | 0.782 (0.085) | 0.434 | 0.000 |
| Random Forest with Hyperparameter Tuning | 0.932 (0.041) | 0.808 (0.075) | 0.809 (0.074) | 0.461 | 0.000 |

## 4. CONCLUSION

Our study explored how different activities—watching a movie clip, listening to music before walking, and listening to music while walking—affect emotion recognition using wearable technology. Building on the work of Juan C. Quiroz, PhD, we introduced Random Forest with Hyperparameter Tuning, significantly enhancing accuracy compared to the baseline model, Logistic Regression, and the default Random Forest. This approach achieved mean accuracies of 86.63% for happy vs. sad emotions and 76.33% for happy vs. neutral vs. sad emotions, demonstrating the effectiveness of fine-tuning model parameters. Despite promising results, our study's limited sample size and controlled conditions may affect generalizability, highlighting the need for future research with larger, more diverse datasets and real-world settings. Our findings emphasize the potential of advanced machine learning to improve emotion recognition, with significant implications for mental health monitoring and human-computer interaction, contributing to ongoing advancements in this field.

## ACKNOWLEDGMENT

Special thanks to the School of Computing, Telkom University, for providing the resources and environment that facilitated this work. The support from the faculty and access to state-of-the-art facilities were instrumental in the successful completion of this research.